\documentclass[11pt]{article}
\usepackage{moriond,epsfig}

\usepackage{epstopdf}

\def\be{\begin{equation}}
\def\ee{\end{equation}}
\def\bea{\begin{eqnarray}}
\def\eea{\end{eqnarray}}

\begin{document}
\begin{flushright}LYCEN-2009-07\end{flushright}
\vspace*{4cm}

\title{DIRECT DARK MATTER SEARCHES AND THE EDELWEISS-II EXPERIMENT}

\author{ J. GASCON }

\address{IPNL, Universit\'e de Lyon, Universit\'e Lyon 1, 
                 CNRS/IN2P3, 4 rue E. Fermi, 69622 Villeurbanne, France}

\maketitle\abstracts{
Direct searches for Dark Matter are experiment dedicated to
the observation of the energetic recoiling ions produced by the scattering
of WIMP particles from our galactic halo on terrestrial targets. 
The status and prospects of some currently running  experiments are presented,
together with new preliminary results of the experiments EDELWEISS-II.
}

\section{Introduction}

Direct searches of Dark Matter\cite{rev} consist in experiments dedicated to
the observation of the recoiling ions produced by the scattering
of Dark Matter particles from our galactic halo on terrestrial targets. 
This endeavour is motivated by the gathering of evidence for
the presence of cold Dark Matter at all scales of the Universe
(as in, e.g. Ref.\cite{wmap}).
In addition, many of the supersymmetric models that are soon to be tested at
the LHC offer a valid candidate for these Weakly Interacting Massive Particles
(WIMP): the neutralino.

Cosmological observations have revealed the presence of Dark Matter,
and continue to provide increasing details to its distribution, past and present, 
in the Universe.
However, it is only by producing these new particles at colliders 
that it will be possible to ascertain their true nature.
The observation of WIMP scattering on a terrestrial target 
at the expected rate would be crucial for establishing a formal
link between cosmological Dark Matter and a given particle.
Another crucial test would be the observation in cosmic rays of remnants
from WIMP annihilations.
This is the subject of indirect searches,
which are covered by other presentations at this conference.
These two types of searches are complementary.
For instance, in indirect searches, the signal strength depend on both the
scattering and the annihilation cross-sections of WIMPs,
as well as the square of the WIMP density.
In direct searches, the rate is a linear function of the WIMP density and,
for a given density, it depends on the scattering cross-section only.

Currently running experiments are sensitive to rates of the order
of one WIMP interaction per month and per kg of target.
In terms of a spin-independent coherent scattering on a nucleon,
this correspond to a scattering rate of 5$\times$10$^{-8}$pb,
approaching the supersymmetric model predictions in 
the so-called "Focus Point" region.
A more comprehensive coverage of supersymmetric model
predictions requires sensitivity to cross-section of the order
of 10$^{-9}$ to 10$^{-10}$ pb.
This corresponds to
rates of the order of one interaction per year and per ton,
which is the challenging goal of future projects.

\section{Principles of Direct Dark Matter Searches}

The relevant scale for direct Dark Matter searches
can be set using general arguments.
First, the velocity of the WIMP particles in our
vicinity must be similar to the typical
velocity of any other object trapped
in the gravitational well of our Galaxy,
{\em i.e.} approximately 200 km/s.
An appropriate mass scale for the WIMP is that at which
supersymmetric particle are expected and sought for:
$\sim$100 GeV within one order of magnitude.
From these two numbers we can conclude that the
elastic collision of a WIMP with a nucleus should
give it a kinetic energy ot the order of 20 keV.
This average does not vary significantly when more realistic 
velocity distributions are considered\cite{lewin}.
More precise predictions require a better understanding of
Dark Matter halo shapes, density profiles and velocity distributions. 
The question of their homogeneity could
have significant impact on the observable scattering rate.

The event rate depends on the local WIMP density
and on the scattering cross-section.
The actual shape of the Dark Matter halo is the subject of intense
debate, especially for its most inner part. Luckily, the sun lies
in the outer part of the visible disk of our Galaxy, where there is
less uncertainty on the WIMP density.
A value of 0.3 GeVcm$^{-3}$ is generally adopted,
at least in models where inhomogeneities do not play a
significant role.
The cross-section predictions are model-dependent.
In most models, the scattering cross-section of a WIMP
on a nucleus is dominated by the coherent
spin-independent interaction on all nucleons.
In order to compare their relative sensitivities, most experiments
express their rates, expressed as events per unit of detector mass 
and exposure time, in terms of the spin-independent 
WIMP scattering cross-section on a single nucleon,
using the standard prescriptions of Lewin and Smith\cite{lewin}.
These cross-sections can also be compared with model predictions, 
although this operation is more model dependent than detector-to-detector
comparisons.
Current supersymmetric model predictions for the WIMP-nucleon
cross-section are typically in the range from 10$^{-8}$ to 10$^{-10}$~pb.
The range has evolved
in time as new observations in particle physics and cosmology
have brought additional constraints on supersymmetric parameters.
Direct searches are also starting to constrain it, with CDMS\cite{cdms} and
XENON\cite{xenon} having reached recently sensitivites of the order
of 5$\times$10$^{-8}$~pb.

The true challenge of direct searches doesn't lie in the relative
uncertainty of the predicted rate, but rather in its extremely low value. 
As an example, a of one event per kg per year is then order of
magnitude below the natural activity in the the human body.
Modern neutrino experiments are accustomed to interaction rates 
of a few events per day and per kiloton of target material, 
but in this case the energy scale of interactions is the MeV.
In dark matter searches, the energy released is one hundred times less,
and can be comparable to X rays from relatively light materials.
Dark Matter searches are probing a domain of ultra-low
radioactivity in an energy domain that has never been probed before. 
It can do it successfully by using powerful discrimination techniques,
but the required high rejection rates require a detailed understanding
of the tails of the distributions of the discriminating variables,
and an excellent detector reliability.

Direct Search detectors must be build with materials with
extremely low radioactivity, and protected from the
ambient background by efficient shields.
Cosmic activation is reduced by installing the experiments in
deep-underground sites.
Nevertheless, these measures are not sufficient to reduce
the background to the required levels, and an active rejection
of background is required.
The most efficient discrimination strategy is to identify the nature
of the recoiling particle. In the case of a WIMP scattering, it is
a heavy ion, often called "nuclear recoil".
The bulk of the natural radioactivity involves electron recoils, 
either due to Compton or photoelectric gamma-ray
interactions, or to $\beta$ rays.
Nuclear recoils are stopped in less than 20 nm in a
solid substrate. It is thus very difficult to extract information
on the direction of ion recoils, and for this reason no large detector
has yet the sensitivity to detect the directionality of the WIMP flux
due to the sun velocity. 
However the linear energy loss of a heavy ion is significantly more important
than the value for an electron of the same energy.
This feature is exploited in detectors like PICASSO\cite{picasso}
and COUPP\cite{coupp}, where this large energy density is
used to trigger bubble formation in superheated liquids.

Another significant difference between nuclear and electron recoils
is the relative ionization and scintillation yields associated with the interaction.
For example, a ion recoiling in a crystal will dissipate most of its energy
in lattice deformations and vibrations, resulting into phonon excitations
that will quickly decay into a thermal equilibrium.
A recoiling electron interacts directly with the other electrons and has
a larger ionization yield. The relative ionization yield of the two processes
is well described by the Lindhard theory\cite{lindhard}.
In germanium, the ratio of yields is three to four, depending on the energy range.
Scintillation yields can vary by even larger amounts\cite{qscint}.
Many experiments are developing detectors where the discrimination
is based on the comparison of two signals, such as ionization and
scintillation\cite{xenon,zeplin}, scintillation and heat\cite{cresst}
or ionization and heat\cite{cdms,edelweiss}

After the nuclear recoil identification is applied, a remaining background is 
the one associated with the elastic scattering of fast neutrons on target nuclei.
The neutron flux on the detector must be moderated with a thick low-A shield.
The most sensitive experiments must also contend with the flux of neutron
due to the interactions in the detector support and the gamma-ray
shielding of the few cosmic rays that can reach the underground site.
The experiments are surrounded by an active veto
that can tag these muons. 
Neutron interactions can also be tagged by their short ($\sim$cm)
mean free-path in solids. 
As a consequence, neutrons tend to be associated
with surface events in large-volume detectors, and with
multiple-hit events in segmented detector arrays.
However, the similarity of the detector response to 
neutron and WIMP interactions is very useful for
a precise on-site calibration of the detector.

Given the uncertainties associated with the ultra-low
background required for these experiments, it is essential
to develop detectors with more than one type of target nucleus
in order to exploit the $A^2$-dependence of the cross-section
which is expected for a coherent spin-independent scattering.
Targets with nuclear spins are of interest to investigate the
special case where the spin-dependent cross-section dominates
despite the $A^2$-scaling of the spin-independent one.
With clean samples of more than 10$^{4}$ WIMPs interactions,
it may also be worth to look for seasonal variations of the rate,
although the actual size and phase of the modulation
depends on the details of the halo model
(see e.g. Ref.\cite{stream}).

From the $A^2$-dependence of the coherent cross-section, one
could conclude that the detectors with the largest-$A$ target are favoured. 
This is not systematically the case. This dependence
is partially offset by nuclear form factor effects\cite{lewin}.
More importantly, the observed rate has a strong dependence
on the achieved experimental threshold. 
The currently-running experiments with the best sensitivities
use a variety of targets and techniques.
The largest progress usually comes from the improved understanding 
of backgrounds, the tight control  of detector imperfections and 
the steady technological advances to get rid of both of them.
This favours technologies offering enough precision and versatility
for studying in details the present-day backgrounds and detector
imperfections.

\section{Currently running experiments}

\subsection{Experimental limits}

At present, the best experimental limits for spin-independent 
scattering cross-section
for WIMPs masses above 40 GeVc$^{-2}$ are those of the CDMS 
experiment\cite{cdms}.
At lower masses, the experiment XENON\cite{xenon} provides the best limits.
In both cases, the best limits are at the level of 5$\times$10$^{-8}$pb 
at 90\%CL.
The limits are above 10$^{-7}$pb in the WIMP mass range 
from 15 to 1000 GeVc$^{-2}$.
The sensitivity  degradation at low masses is due to the 
experimental thresholds.
At high masses, it is caused by the asymmetry between the target
and projectile masses.
These limits are two order of magnitude below the interpretation of the
DAMA oscillation, discussed in this conference,
in terms of WIMP using Lewin and Smith prescriptions.
Reconciling this oscillation with the limits from the searches
requires a significant departure from these prescriptions and
has prompted a wealth of alternate specific models.
Solutions involving WIMPs with dominant spin-dependent interactions
on protons are at odds with the limits from the experiments COUPP\cite{coupp}
and KIMS\cite{kims}.
WIMPs with masses below the XENON and CDMS limits are excluded
by the high-resolution germanium experiment CoGeNT\cite{cogent}

These alternative models widen the choice of detector technologies and 
experimental strategies.
However, the best sensitivities to 
WIMPs corresponding to the more standard mSUGRA
models that will be tested at LHC are attained by detectors
aiming at coherent spin-independent interactions.
In the following, we will briefly describe the two most
sensitive experiments at present, CDMS and XENON, 
and present the recent progress of the EDELWEISS  with 
germanium detectors
demonstrating an unprecedented rejection of events associated
to surface contaminations.

\subsection{XENON two-phase detector}

As mentioned before, the discrimination of electron and nuclear recoils
of this type detector is based on the difference in the relative ionization and
scintillation yields.
Rare gases are a prized target material, as a very high-radiopurity can be
achieved by multiple purification cycles.
In rare gases, the scintillation comes from the de-excitation of a meta-stable
excimer, efficiently produced in atomic collisions.
Ionization is less efficient at producing this excimer.
The principle of the XENON-10 detector\cite{xenon} is the following:
a cylindrical cell is filled with 10 kg of liquid xenon, 
viewed by two arrays of $\sim$45
photomultipliers, one at the bottom and one at the top.
The first signal comes from the scintillation observed directly following the
interaction.
An intense electric field drifts the electrons toward the top surface.
They are further accelerated in the gaseous phase above the liquid,
creating a secondary pulse of scintillation proportional to the
ionization yield.
The ratio of the secondary signal to the primary signal varies by
a factor $\sim 0.4$ depending on the nature of the recoil.

In addition to this ratio, the detector provides full 3-dimensional coordinates
of the primary interaction. The position along the vertical axis is given
by the time delay between the two pulses, and the position along the plane
is deduced from the signal intensity in the different photomultipliers.
As most interactions due to the radioactive background occur near
the surface of the xenon volume, only those occurring in the inner 
fiducial volume of 5.4 kg are considered.
The limits reported by XENON are based on a
total of 10 events observed in 59 days in this fiducial volume.
The discrimination performances are limited by the statistics of
scintillation photons.
The observed events can be interpreted as the effect
of the pile-up of Compton interactions, with one of them occurring
outside the active volume for charge collection.
This imperfection should be cured in the next generation of the experiment,
XENON-100, currently being commissioned in the Gran Sasso
National underground Laboratory.
Another improvement should be an increase of the light collection,
to get a better separation of nuclear and electron recoils. 

Another two-phase experiment using xenon is ZEPLIN-III\cite{zeplin}.
Alternatively, other experiments, like ArDM\cite{ardm}, 
WArP\cite{warp},  and DEAP/CLEAN\cite{deap}
plan to use argon instead of xenon as a target.
The lower atomic mass is offset by the use of a cheaper and more
readily available material,
and the possibility to identify nuclear recoils
by using a pulse shape discrimination exploiting the long
($\sim \mu$s) lifetime of one of the excimer states.
This technology must contend with the important radioactive 
background from the naturally occurring radioisotope $^{39}$Ar
(10$^5$ decay per kg per day).

\subsection{CDMS}

The experiment CDMS has obtained its limits on WIMP interaction
using an array of fifteen 250~g cryogenic germanium detector, 
with the simultaneous measurement of phonon and ionization signals.
These detectors have two natural advantages: both phonon and
ionization measurements have typical energy resolutions
of one keV or less, and the radioactivity of hyperpure semiconductor  
crystals is extremely low. 
The response to nuclear recoils for this material
is known in detail\cite{qchal}.

The ionization signal is collected using electrodes covering the surface 
of the crystal.
The phonon signal is measured using an array of four quadrants of 
$\sim$1000 transition-edge tungsten sensors,
covering one face of the detector.
This system has a large acceptance to out-of-equilibrium phonons.
As a result, the relative size of the phonon signal in the four quadrants
depends on the position of the interaction 
in the plane parallel to the sensor. 
Using the amplitudes recorded in the four  quadrants, 
it is possible to reconstruct both this location, with a mm
precision, and the energy, with a keV resolution.
What is gained by the sensitivity to athermal phonons is an event-by-event
information of the time structure of the build-up of the signal.
The phonon signal rise time and its delay relative to the fast ionisation
signal are of the order of a few $\mu$s.
It is observed that these time constants are systematically larger for
phonons associated with nuclear recoils, relative to those arising from
electron movements\cite{cdms}.

These detectors thus provide two independent discrimination variables for
the identification of nuclear recoil: the ionization yield relative
to the total calorimetric energy measured by the phonon sensor,
and the time structure of the phonon signal.
This double discrimination is important, since the ionization yield
measurement is degraded for surface events, where a substantial
fraction of the charge may be lost due to trapping and diffusion effects.
With this discrimination, CDMS was able to accumulate a fiducial 
exposure of 121 kgd
with no events observed in the region where nuclear recoils are expected.
The background due to surface events with bad charge collection for 
that experiment was estimated to be 0.6$\pm$0.3 events.
More data is being recorded and analysed now, and with the development
of more efficient background cuts, the experiment should reach a sensitvity
to spin-independent cross-sections of 2$\times$10$^{-8}$pb by the end of
2009.
The collaboration is preparing for the next generation of arrays with total
germanium masses of 40~kg to 190~kg (superCMDS) 
and an eventual one-ton stage (GEODM).

\subsection{EDELWEISS}

The experiment EDELWEISS uses 350~g germanium cryogenic
detectors,  installed in the Laboratoire Souterrain de Modane (LSM)
in the  Frejus Tunnel between Italy and France.
This deep underground site is well suited for experiments
aiming at sensitivities well below 10$^{-8}$pb.
As CDMS, its detectors use the ratio of the ionization yield
to the phonon signal to identify nuclear recoils.
Here, the phonon measurement is provided by a simple 
GeNTD thermistance, glued to the detector.
The signal is purely thermal, with a uniform response over the
entire detector volume.
In contrast with CDMS, the rejection of events with incomplete charge
collection is not based on the phonon signal, but on ionization. 
The electrodes on the flat surfaces of the cylindrical detectors
are replaced by concentric, annular interleaved electrodes,
with a pitch of 2~mm.
With this "InterDigit" electrode design (ID)\cite{id},
surface events are tagged by the presence of charge on
two electrodes on the same side of the detector.
The phonon measurement is provided by a simple 
GeNTD thermistance, glued to the detector.

\begin{figure}
\begin{center}
\psfig{figure=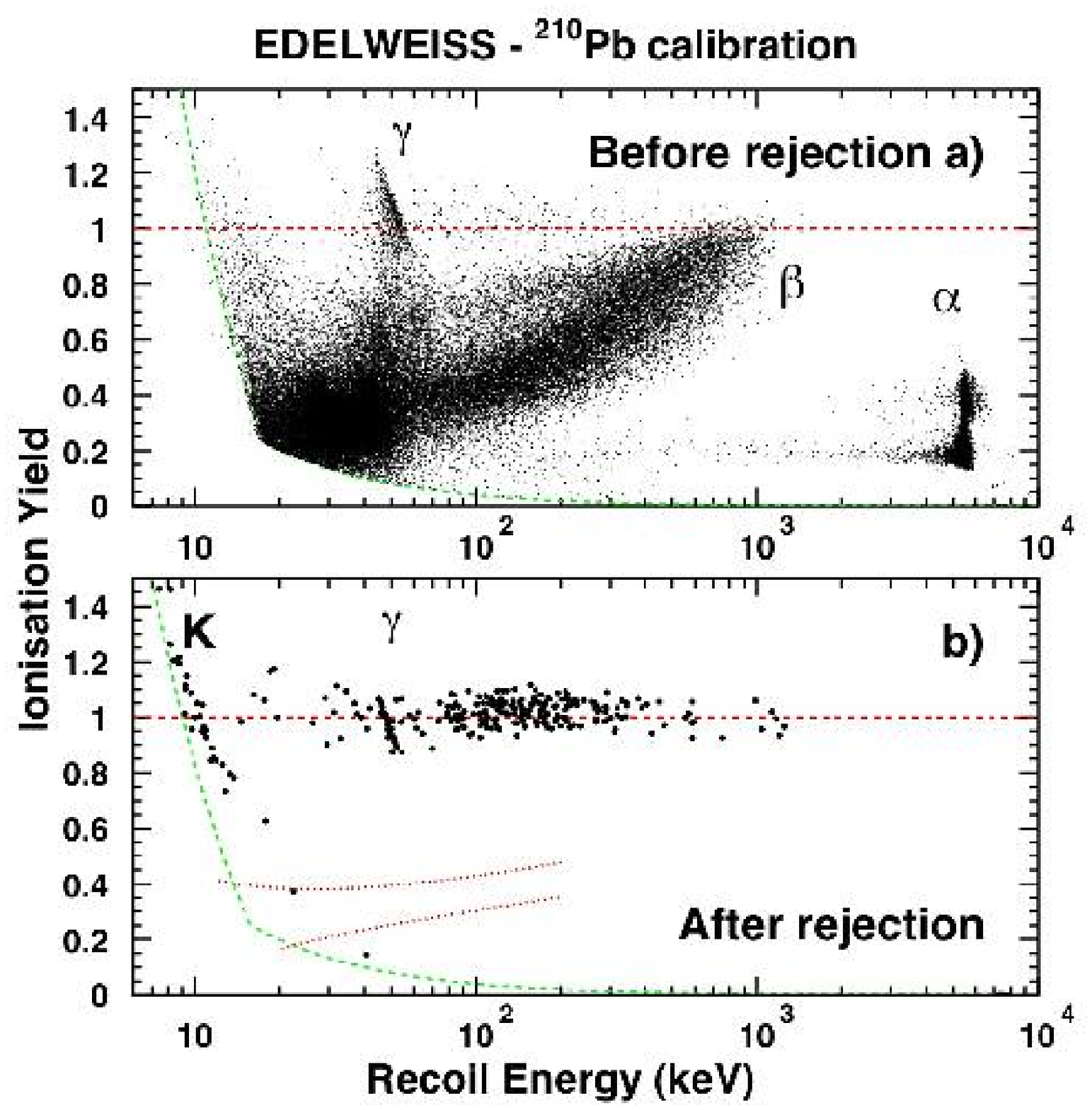,height=7.5cm}
\psfig{figure=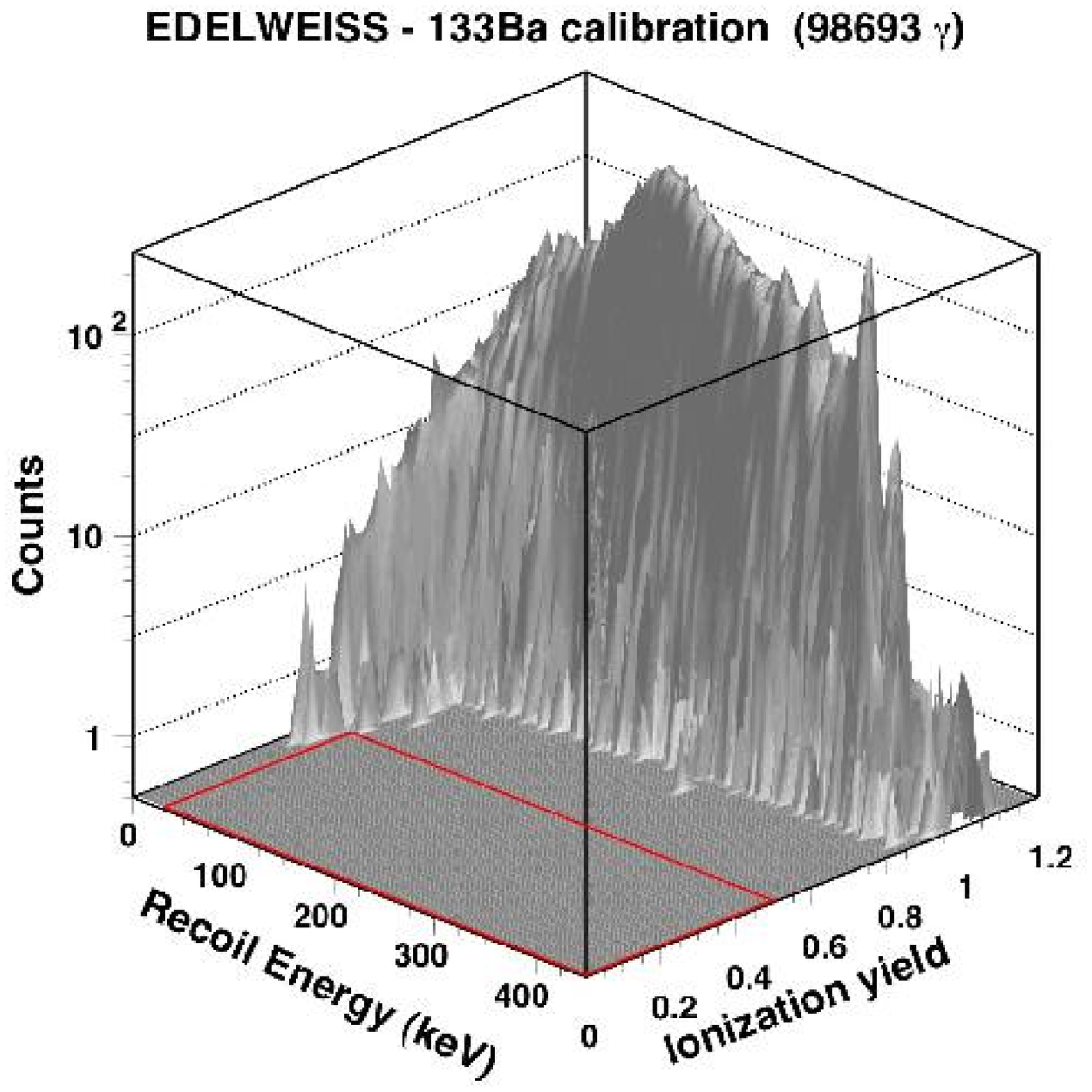,height=7.5cm}
\caption{Ionization yield as a function of recoil energy recorded in EDELWEISS
InterDigit detectors (ID) recorded in calibrations with a $^{210}$Pb (left) and
$^{133}$Ba (right) source (taken from Ref.~\protect\cite{id}). 
The pannels (a) and (b) show the effect of surface
event rejection on the population of $\gamma$, $\beta$ and $\alpha$ rays
from the source (see text for explanation).
The remaining $\gamma$  and K-shell X-ray 
 population along the ionization yield of one
in pannel (b) comes from natural background.
The right pannel show the rejection for $\gamma$-rays.
\label{fig-idrej}}
\end{center}
\end{figure}

This method is very efficient to reject surface interaction,
as shown in Fig.~\ref{fig-idrej}a and b, where the ionization yield
of events recorded in a detector exposed to a $^{210}$Pb
source are shown as a function energy.
The data of Fig.~\ref{fig-idrej}a is dominated by  $\beta$ and $\alpha$ rays. 
When the rejection criteria are applied, only one $\beta$ event remains
in the region where nuclear recoils are expected (dashed lines
around ionization yields of $\sim$0.3 in Fig.~\ref{fig-idrej}b).
This corresponds to a rejection factor of $\sim$10$^{5}$ for this type
of events. This is amply sufficient to reject the observed surface
contamination in EDELWEISS, measured at rate of a few
events per day per detector\cite{fiorucci}.
Fig.~\ref{fig-idrej}c shows that the rejection for $\gamma$-ray events 
can reach the value of 10$^{4}$ that is required for reaching
a sensitivity of 10$^{8}$pb.

\begin{figure}
\begin{center}
\psfig{figure=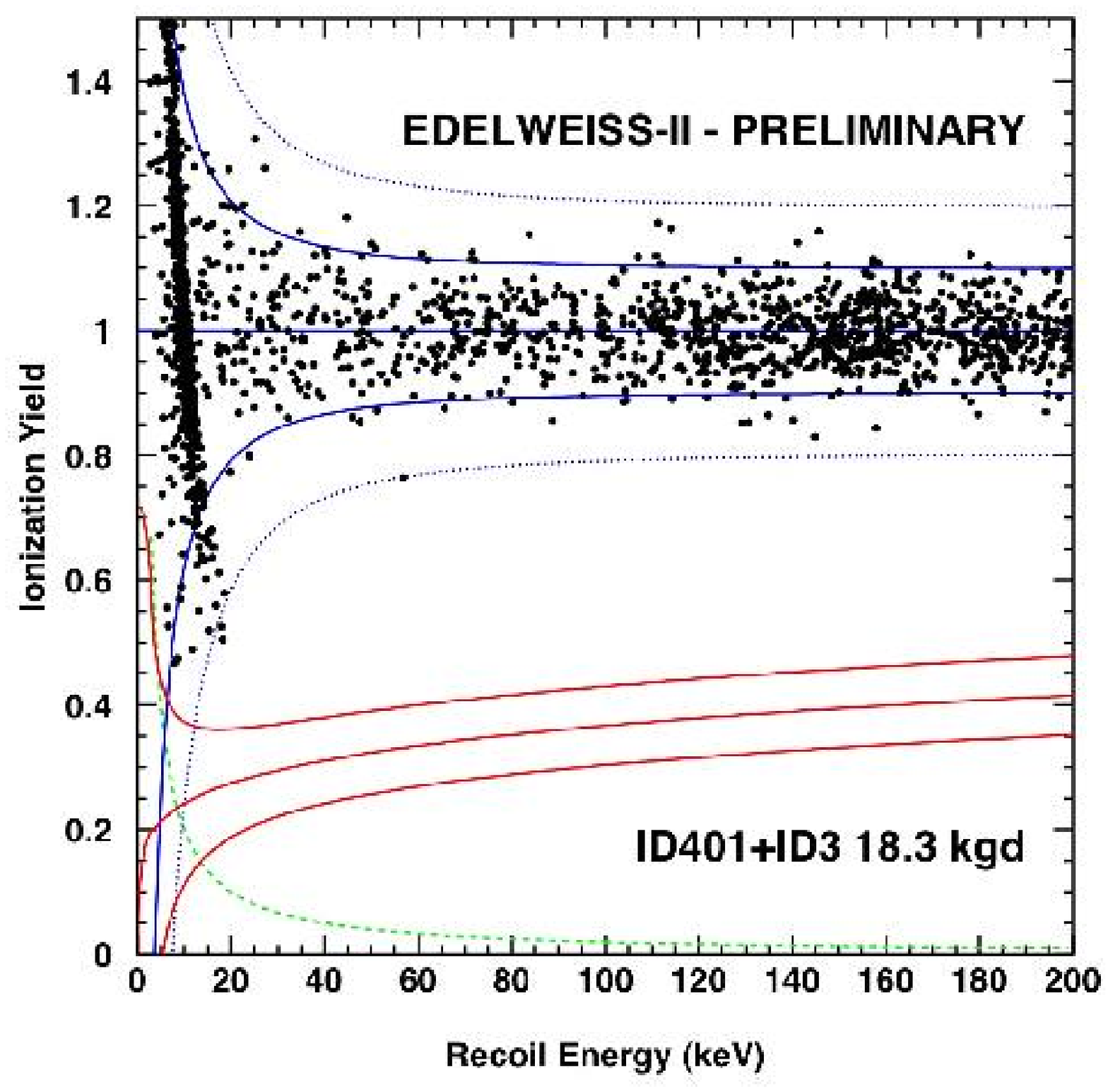,height=7.5cm}
\psfig{figure=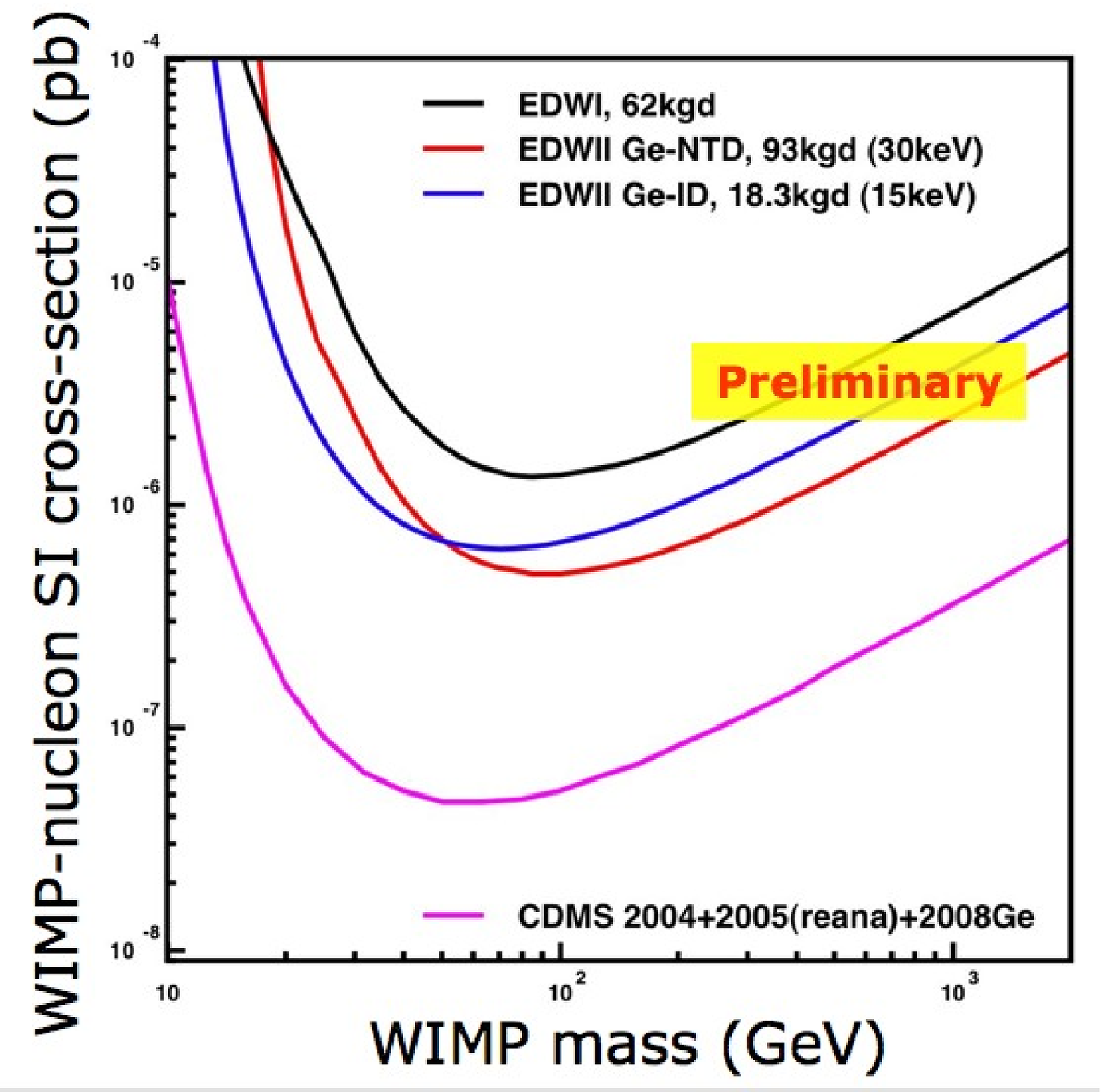,height=7.3cm}
\caption{Left: Ionization yield as a function of recoil 
energy recorded in EDELWEISS
InterDigit detectors (ID) recorded in and exposure of 18.6 kgd. 
Right: 90\% CL limits for spin-independent scattering  cross-section for WIMPs
 as a function of the WIMP mass. The curves are (from top to bottom,
 above 100 GeV): the 2005 results from EDELWEISS in its
 previous setup; the result of the 18.3 kgd recorded with
 detectors with ID electrode design; 
 the 2008 results in the new setup with 93 kgd 
 without the ID electrode design and the CDMS combined limit.
\label{fig-iddata}}
\end{center}
\end{figure}

These ID detectors were designed, build and tested in 2007 and 2008.
The EDELWEISS collaboration is now operating ten 400~g ID detectors
in its low-background facility at the LSM.
The radioactive backgrounds in this setup has been thouroughly
studied in 2007 and 2008 using detectors without the interleaved design.
They have been reduced relative to the previous setup of the
experiment\cite{edelweiss}. 
The muon veto has been commissionned, and is observing coincidences
with the germanium array at the levels of a few events per 100 kgd.
In 2008, an fiducial exposure of 18.3 kgd was recorded with the
first two 400~g ID detectors (Fig.~\ref{fig-iddata}a).
No nuclear recoils were observed, down to a threshold of 10 keV
in recoil energy.
The efficiency for nuclear recoils reaches its plateau below a
recoil energy of 15 keV.

This data was interpreted in terms of limits for spin-independent scattering 
cross-section for WIMPs, as a function of their mass (Fig.~\ref{fig-iddata}b).
This limit is comparable to what has been obtained in an exposure
of  93.6 kgd of detectors without the ID design,
despite the five-fold difference in exposure.
This shows the importance of surface event rejection, as the
limit derived from the larger exposure is constrained by the
appearance of three nuclear recoil candidates in the range from 30
to 45 keV.
With the increase of number of ID detectors (from two to ten)
and of the exposure, EDELWEISS should reach a sensitivity
to $\sim$5$\times$10$^{-8}$ by the end of 2009.
Further improvements should come with the addition of new detectors,
and by the development of detectors with an increased fiducial
volume.
The ID technology has proved to be a simple and efficient way to
address the problem of surface events, and current studies
of muon-induced events seem to suggest that the present
EDELWEISS setup could be suitable to probe cross-sections
of the order of 10$^{-9}$pb.
Further developments are being studied in the framework of
the EURECA\cite{eureca} collaboration.

\section{Prospects and conclusions}

In the domain of direct Dark Matter searches, there is an intense
competition between detectors with different target nuclei and
detector technologies.
This is a welcome diversity since the observation of a WIMP
signal will necessarily require confirmation, ideally in a
detector with a target nucleus with a different atomic mass.
At present, the germanium cryogenic detectors of CDMS
and two-phase detector of the XENON collaboration
have achieved the best sensitivities.
Present limits are just starting to probe the 10$^{-8}$pb range
where an important class of supersymmetric models relevant
for LHC are lying.
Both collaborations are developing more ambitious projects
aiming at cross-sections in the range of 10$^{-9}$pb and
eventually 10$^{-10}$pb.
In Europe, the efforts for cryogenic detectors have been
federated in the collaboration EURECA, aiming at deploying
a ton-size array of heat-and-scintillation and heat-and-ionization
detectors in the future extension of the LSM laboratory.

It is not yet clear what detector technology will take the lead
in the coming years.
The present results shown that two-phase detectors and cryogenic 
germanium arrays are serious contenders.
Two-phase detectors offer large masses, but the problem
of light yield must be closely addressed, as it affects the
discrimination capabilities. 
The problem associated with long drift lengths requires
also attention.
Cryogenic detectors  have a better resolution and their
rejection is superior, but scaling up raises the questions of
price and optimal detector size.
One should not exclude that other technologies may catch 
up and surpass present-day sensitivities.
This competition is however vital for the field, as the
formal identification of the WIMP as a new particle in large abundance
in our environment will be a small revolution that will
call for extensive experimental verification.

\section*{Acknowledgments}
The author wishes to thank the organizers of this wonderful conference,
and the EDELWEISS Collaboration for making available their preliminary data
for this review.


\section*{References}
\begin{thebibliography}{99}

\bibitem{rev} G. Jungman, M. Kamionkowski, and K. Griest, 
       Phys. Rep. \textbf{267}, 195 (1996); 
G. Bertone, D. Hooper and J. Silk, Phys. Rep. \textbf{405}, 279 (2005).

\bibitem{wmap} E. Komatsu, et al., Astrophys. J. Suppl. {\bf 180} (2009) 330.

\bibitem{lewin}J.D. Lewin and P.F. Smith, 
      Astropart. Phys. \textbf{6} (1996) 87.

\bibitem{cdms} Z. Ahmed et al., Phys. Rev. Lett. \textbf{102} (2009) 011301.

\bibitem{xenon} J. Angle et al., Phys. Rev. Lett. \textbf{100} (2008) 21303.

\bibitem{picasso} F. Aubin et al., New J. Phys. \textbf{10} (2008) 103017. 

\bibitem{coupp} E. Behnke et al., Science \textbf{319} (2008) 933. 

\bibitem{lindhard}J. Lindhard \emph{et al,} 
    K. Dan. Viderask. Selsk., Math. Fys. Medd
    \textbf{33} (1963) 10 and \textbf{36} (1968) 10.

\bibitem{qscint} I. Bavykin et al., 
     Astroparticle Physics \textbf{28} (2007) 489.

\bibitem{zeplin}  V. N. Lebedenko et al., arXiv:0812.1150 [astro-ph]. 

\bibitem{cresst} G. Angloher et al., 
     Astroparticle Physics \textbf{31} (2009) 270.

\bibitem{edelweiss} V. Sanglard et al., 
     Phys. Rev. D \textbf{71} (2005) 122002 .

\bibitem{stream} C. Savage, K. Freese and P. Gondolo, 
     Phys. Rev. D \textbf{74} (2006) 043531. 

\bibitem{kims} H.S. Lee et al., Phys. Rev. Lett. \textbf{99} (2007) 091301.

\bibitem{cogent} C. E. Aalseth et al.,
     Phys. Rev. Lett.  \textbf{101} (2008) 251301.

\bibitem{ardm} L. Kaufmann and A. Rubbia, 
     J. Phys. Conf. Ser. \textbf{60} (2007) 264.

\bibitem{warp} P. Benetti et al., 
     Nucl. Instrum. Meth. A \textbf{574} (2007) 83.

\bibitem{deap} M.G. Boulay and A. Hime, 
     Astropart. Physics \textbf{25} (2006) 179.

\bibitem{qchal} A. Benoit et al., 
     Nucl, Instr. Meth. in Phys. Res.  \textbf{A 577} (2007) 558.

\bibitem{id} A. Broniatowski et al., arXiv:0905.0753[astro-ph]. 

\bibitem{fiorucci} S. Fiorucci et al., Astropart. Phys \textbf{28} (2007) 143.

\bibitem{eureca} H. Kraus et al., J. Phys. Conf. Ser. \textbf{39} (2006) 139.

\end{thebibliography}
\end{document}